\documentclass[11pt, a4paper]{article}

\begin{document}
\title
{Dilational viscosity of Langmuir monolayers}
\author{Vladimir Kolevzon\\
 Institute of Physical Chemistry, University of Karlsruhe,\\
76128 Karlsruhe, Germany}
\maketitle
\noindent
The dilational viscosity $\epsilon'$ of the Langmuir monolayer is
considered in a theoretical model taking into
account an orientational effect of the dilational wave on the 
surface molecules. The surface tension treated in the framework of
this model is supposed to be dependent on the degree of molecular
ordering in the monolayer plane. This orientational order is
described by the surface order parameter Q and the
orientational part of the free energy which is given by Landau's
expansion in powers of Q. 
The magnitude of surface 
viscosity determined by the surface tension derivative $\partial
\gamma/\partial Q$ is found to be in good accord with the 
experimentally observed $\epsilon'$. The sign of $\epsilon'$ 
is positive which indicates that increased ordering in 
the monolayer plane increases the surface tension.\\


Strong damping of capillary waves on the monolayer--water interface
has been well known for many years. However the physical nature of the two surface 
viscosities which are included into a dispersion relation \cite{kram, lang}
describing the propagation of surface waves still remains unclear. 
It has been suggested
that the monolayer possesses up to three surface
viscosities: dilational, shear in the interfacial plane and
shear normal to the surface plane \cite{kram,good}. 
Only two of them: dilational $\epsilon'$ and transverse
shear $\gamma'$ do couple to surface modes which scatter
light \cite{lang}. These surface viscosities were introduced as
the response functions which appear after expansion of the
complex moduli $\gamma$, $\epsilon$, to linear order in
$\omega$ \cite{kram}. Hence the dilational modulus is usually 
written as a superposition of the dilational surface viscosity
$\epsilon'$ and the surface elasticity $\epsilon_0$ \cite{kram,good}:
\begin{equation}
\label{lin}
\epsilon=\epsilon_0+i\omega\epsilon'
\end{equation}
The surface viscosity $\epsilon'$ so introduced is a pure phenomenological 
property which was suggested as the surface excess
quantity \cite{good,baus}. 

Only for two-component systems of soluble surfactants is there 
a strong theoretical basis for the dilational surface viscosity
\cite{levi,luka}. In such a system the
surface viscoelasticity appears as a result of 
the competition between adsorption and
diffusion in the concentrational boundary layer beneath the
surface. The main result of this formalism is 
that $\epsilon'$ is not an independent property but is a function
of some relaxation time $\tau$ which depends on the diffusion 
coefficient D and the slope of the adsorption isotherm
$\frac{dC}{d\Gamma_s}$ \cite{levi,luka}: 
\begin{equation}
\epsilon'=\frac{\epsilon_0\tau}{\omega(1+2\tau+2\tau^2)}
\end{equation}
where $\tau=\frac{dC}{d\Gamma_s}\sqrt{D/\omega}$,  
 (C is the bulk concentration and $\Gamma_s$ the surface concentration). 
However for an insoluble (spread) monolayer on a liquid substrate there is
no adequate theoretical description for the dilational surface
viscosity. At the same time
the experimental findings  of negative $\epsilon'$ 
in some liquid systems \cite{earn1, kol} indicate that some
important physical processes are not accounted for in the theoretical
formalism of \cite{kram,luka}.

Our previous papers concerned the dilational viscosity of
various liquid surfaces \cite{kol2,kol3}. It was demonstrated 
that the surface dilational modulus can be described by relaxation of 
fluctuations of temperature and surface charge density. 
Unfortunately the dilational viscosity calculated from those
models appears to be a few orders of magnitude smaller than that 
observed in experiments on different liquid systems 
\cite{earn1, kol}. 
The present paper demonstrates that the orientational effect of
the surface wave on long molecules can be the reason for the
dilational viscosity of the liquid-monolayer interface.

The orientational effect of a shear flow on long molecules
in the liquid bulk has been known for many
years \cite{fren}. However such an orientation of molecules
in the surface layer has not yet been reported.
We concentrate upon a monolayer being in a liquid-expanded (LE) 
state where the surface area per molecule is relatively large
according to the surface isotherm \cite{adam}. Due to this 
fact and strong electrostatic interaction between the molecular head 
groups and water the long axes of molecules are not oriented along
the surface normal but comprise some angle to it.

Unfortunately
LE state is not too much studied experimentally compared to 
a condensed state. Therefore we will use the results of theoretical
predictions on order parameters in LE state. One of them
 \cite{halp} develops a concept of orientational order  
 in a monolayer comprising hard rods grafted on the surface. The main result
 is that molecular alignment increases  continuously with increasing
 surface concentration.
Since the molecular azimuths are not arbitrary but are defined by
the tilt angle, the molecular alignment in the surface plane is
induced by collective tilt. A numerical analysis of flexible molecules
\cite{shik} shows that in-plane ordering in a liquid phase
can be even enhanced by chain length. All these papers presume that
surface fluctuations destroy the in-plane long range order. We will give
a highlight to the effect of dilational waves on the surface order.
As it will be shown below a very small magnitude of the order parameter
is sufficient for our model. 

As mentioned in \cite{halp}, a monolayer formed by
amphiphile diblock copolymer on water surface
can be a realization of the model describing alignment of mobile rods.
Another experimental situation relevant to the present context
is monolayer of a nematogen on water
studied in \cite{halp1}. At low surface concentrations nematic molecules
lie on water but the molecules interact with each other as in a 3-D nematic.

In the present context we suggest that projections of
molecules display a quasi long-range ordering in the surface plane.
In the case of a high tilt angle the short axis of a molecule nearly
coincide with its surface projection.
Let us consider the surface footprint of a tilted molecule 
(see Fig. 1). The long axis of the molecular "cross-section" of 
a length L  makes an angle $\phi$ with wavevector q (supposed to be
along the x-axis) of the dilational wave. This molecule 
experiences an orientation torque induced by the longitudinal 
gradient du$_x$/dx of velocity in the dilational wave as it
was suggested in \cite{prie}. If the center of the
molecule moves with the liquid then the two ends will move
with different relative velocities $u_1$ and $u_2$:
\begin{equation}
u_1-u_2=\delta u=\frac{du_x}{dx}L\cos\phi
\end{equation} 
The transverse (to the molecule axis) component of the relative
velocity gives rise to rotation of the molecule with an
angular velocity $\Omega=\dot{\phi}$
\begin{equation}
\label{omega}
\Omega=-\frac{du_x}{dx}\sin 2\phi
\end{equation}
where negative sign means that the molecule rotates towards to
decrease in $\phi$.

The angular velocity of the molecular rotation is
proportional to the gradient of surface velocity whose magnitude
is small and oscillating: $u \sim u'\exp(i(\omega t-qx))$ .
The effect  is superposed with a strong anisotropy 
of nematic in-plane ordering supposed in the present text.    
The order parameter tensor can be introduced in the
pure 2-D case, analogously to 3-D case \cite{degen}, as :
 $Q_{ij}=Q(n_i n_j -\frac{1}{2}\delta_{ij})$ 
where  i=1,2, n$_i$ is the
nematic director lying in the surface plane. Due to the surface wave 
the order parameter tensor comprises 
two parts: steady Q$^0_{ij}$ and an oscillating one Q$'_{ij}$.

In order to find out the equation describing the behavior of Q$'$
we  use the well-known phenomenological equation \cite{degen}:
\begin{equation}
\frac{dQ_{ij}}{dt}=\lambda s_{ik}-\mu\frac{\partial F}
{\partial Q_{ij}}
\end{equation}
where F is the free energy density, $\lambda$ is a 
proportionality coefficient, s$_{ik}$
is the tensor of viscous stresses and $\mu$ is some coefficient having
the dimension of viscosity. 
Usually the free energy is expanded in 
powers of Q and the expansion contains the terms proportional to
the second, third and fourth power of Q  (see for
example \cite{degen}). The main feature of the 2-D case is an absence of a cubic term
in the energy expansion due to the symmetry in the 
surface plane. Thus the free energy density is written as: 
\begin{equation}
\label{energ}
F =A Q_{ij}Q_{ij } +B Q_{ij}Q_{ij}Q_{kl}Q_{kl}
\end{equation}
where A is the temperature dependent coefficient changing sign at the point 
of an isotropic-nematic transition and B$>0$,  
and the summation is taken over repeated indexes. In principle, it
might be possible to calculate A and B in the 2--D case as it is done
in the bulk nematic \cite{kats1}. However, an anisotropic
 interaction potential is essentially
unknown to the present study. Therefore the coefficients A and B
will be kept without explicit expressions.
  
Since we have only one non-zero component of the stress tensor 
:$s_{11}=\frac{2}{3}\frac{\partial u_x}{\partial x}$ (see \cite{fren})  
Eq(\theequation) is simplified
\begin{equation}
\label{Q1}
\frac{dQ_{11}}{dt}=\lambda\frac{2}{3}\frac{\partial u_x}{\partial x}
-\frac{1}{\tau}
\frac{\partial f}{\partial Q_{11}}
\end{equation}  
 where f is dimensionless energy density $f=F/F_0/V_0$ and 
$\tau=\mu F_0/V_0$ is some relaxation time. The derivative 
$\partial F/\partial Q_{ij}$ can be found from the substitution of the
averaged plus the disturbed component of the order parameter 
tensor into Eq(\ref{energ}). However we simplify 
the situation suggesting that the molecular axes
are oriented in the surface plane either parallel to the wavevector 
{\bf q} of the dilational wave or perpendicular to it. Then the disturbed component
of the order parameter tensor has a diagonal form, i.e. only two non-zero
components: Q$'_{11}$ and Q$'_{22}$ whereas Q$'_{22}$=-Q$'_{11}=Q'$.
 Thus
the free energy derivative is
\begin{equation}
\frac{\partial f}{\partial Q_{11}}=\frac{\partial}{\partial Q'}
(A(Q_0+Q')^2+B(Q_0+Q')^4)
\end{equation}            
where Q$_0$ is the steady part of the order parameter 
and Q$'$ is the magnitude of an oscillating part. 
Then Eq(\ref{Q1}) can be re-written to the first order in Q$'$
\begin{equation}
\frac{dQ}{dt}=i\omega Q'=-\frac{1}{\tau}Q'(2A +12BQ_0^2)+\lambda
\frac{2}{3}\frac{\partial u_x}{\partial x}.
\end{equation}
$\lambda$ is a phenomenological coefficient such that
the term $\lambda\frac{\partial u_x}{\partial x}=
\langle\Omega\rangle$ expresses the
mean torque exerted on the order parameter by the dilational wave.
Following Ref \cite{stef} this is plausible to suppose 
 the coefficient $\lambda$ is the magnitude of the mean
angular velocity averaged over all molecular orientations:
The averaging gives $\lambda=1$ if  the angular velocity in the form of
Eq(\ref{omega}) is used. Due to its phenomenological origin, $\lambda$ is not
inevitably equal to 1 but may depend of molecular shape,
or diffusion coefficient, therefore only an  approximation $\lambda\approx$1 
will be used. Hence Q$'$ is
\begin{equation}
\label{order}
Q'=\frac{2}{3}\frac{\partial u_x}{\partial x}
\frac{\tau}{i\omega \tau+2A+6BQ_0^2}
\end{equation}
    
The amplitude of Q$'$ is $\omega$ dependent
if the typical wave frequency and relaxation time are such
that: $\omega \sim \tau$. We remind that only the real part of
 the complex Q$'$ is physically meaningful; a phase of Q$'$ 
corresponds to
some angle between the wavevector {\bf q} and the averaged orientation
of the director. The effect of
this phase shift is an additional wave damping 
on the surface possessing some anisotropy.  Our situation 
strongly resembles the propagation of longitudinal acoustic
waves through an anisotropic liquid which was
considered by Frenkel in \cite{fren}. He demonstrated that the
components of the order tensor are proportional to the
longitudinal gradient of velocity in a sound wave. If the wave
frequency is comparable to the characteristic relaxation time 
then the order tensor and the velocity gradient are
connected via the usual Maxwell formula describing the
viscoelastic relaxation of molecules subjected to oscillating 
disturbances.  
Such viscoelasticity points to the existence of a phase shift
between a disturbance and the media response \cite{fren}.

Our goal is to find the surface elastic modulus
which can be written as \cite{luka}:
\begin{equation}
\label{mod}
\epsilon=-\rho_s\frac{\partial\gamma}{\partial\rho_s}=-
\rho_s\frac{\partial\gamma}{\partial Q}\frac{\partial Q}{
\partial\rho_s}
\end{equation}
where $\gamma$ is the surface tension and $\rho_s$ is the
surface concentration. The linkage between disturbances of the
surface concentration and the surface velocity is found from
the conservation of matter at the surface \cite{levi,luka}:
\begin{equation}
\frac{\partial \rho_s}{\partial t} = D_s\frac{\partial^2 \rho_s}
{\partial x^2}-\rho_{s0} \frac{\partial u_x}{\partial x}
\end{equation}
where $\rho_{s0}$ is an equilibrium surface concentration.
We are looking for the surface concentration
in the form of a wave: $\rho_s ' \sim e^{i(\omega t-qx)}$.
Substitution in Eq(\theequation) yields:
\begin{equation}
\rho_{s0} \frac{\partial u_x}{\partial x}=\rho_s'(-i\omega
+D_s q^2) \approx -i\omega\rho'
\end{equation}
where we supposed the surface diffusion coefficient D$_s$ to be nearly
equal to that of the bulk (D $\approx 10^{-5}$
cm$^2$/s) and  typical wave frequencies are such that
$\omega >> D_s q^2$
\cite{lang,levi}. 
It is easy to see that the second derivative in Eq(\ref{mod})
is:
\begin{equation}
\label{firstder}
\frac{\partial Q' }{\partial\rho_s}=-\frac{2}{3}\frac{i\omega\tau}
{\rho_{s0}(i\omega\tau+2A+12BQ_0^2 )}
\end{equation}
where a negative sign means that Q is a decreasing function of
the surface concentration. This is a typical signature of
surface fluctuations;
they tend to destroy a long range nematic order whose order parameter
usually grows with $\rho_s$ in the absence of fluctuations \cite{halp}.


Some papers (\cite{sluck} and references herein) postulate the
 existence of a nematic ordering
induced on the surface of a liquid crystal being in an
isotropic state having Q$\equiv$0. This surface ordering treated in 
\cite{sluck} is due to the
contact with a solid wall inducing a molecular alignment
rapidly decaying into the liquid bulk. In \cite{sluck}
the free surface energy has been written using the Landau
expansion over the in-plane order parameter. Note that liquid
crystals have relatively large penetration depth of the surface
ordering while the ordering in spread monolayers is strictly
confined to the first surface layer. Therefore the 
difference from liquid crystals is an absence of the Frank
elastic modulus describing the reaction of the bulk on the
surface ordering. 

We establish now an analytical dependence
of the surface tension on the surface order parameter.  
For the Langmuir monolayer only the surface has 
some ordering--the liquid bulk is essentially isotropic. 
Therefore the orientational part of the surface 
free energy is given by the integration of F over the liquid 
depth: $F_s=\int_0^\infty F(Q) \delta(z) dz= \xi F(Q_s)$,
where  $\delta(z)$ is the delta function and Q$_s$ is the magnitude of
the order parameter on the surface and $\xi$ is some correlation
length.
 Thus the orientational part of surface free energy density is written as: 
\begin{equation}
\label{ener1}
F_s =\xi k_B T N_A\rho_b/M_m(A Q_s^2 +B Q_s^4)=k_B T N_A\frac{\rho_s}
{M_m} F(Q),
\end{equation}
where M$_m$ is the molecular weight and the relation $\rho_s=\xi\rho_b$
connects the number of particles in the surface and the bulk.
It should be borne in mind that only a minimum of the function
F(Q) corresponds to physically stable state \cite{degen, land}
and minimization with respect to Q gives: $F_{min}$=A Q$_0^2$/2,
where Q$_0^2$=--A/2B \cite{land}. It is apparent that in the 
2-D case we have a continuous phase transition instead
of the first order transition in the bulk liquid crystal. 

 The derivative $\partial\gamma_0/\partial Q$ can be found
from a general thermodynamic equation relating the surface tension 
and the surface excess density of mass $\Gamma_s$ \cite{adam}:
\begin{equation}
d\gamma=- \Gamma_s d\mu_s
\end{equation}
where $\mu_s$ is the surface chemical potential. Note that the entropy term
--S$_s$dT is omitted due to isothermal character of our effects.
The chemical potential is $\mu_s=(\partial F_s/\partial N)_A$, by the
definition \cite{land} and $\Gamma_s \equiv\rho_s$ for any insoluble film.
The derivative $\partial\gamma/\partial Q$ is 
\begin{equation}
\frac{\partial\gamma}{\partial Q}=-\rho_s\frac{\partial\mu_s}{\partial Q}=
-\rho_s k_B T\frac{\partial F(Q)}{\partial Q}
\end{equation}
The variations in the surface mass density upon a phase transition are
negligible.

As we mentioned above only a minimum in the free energy is realized
in the equilibrium. One might think that $\partial F/\partial Q$ is zero
as it should be in the minimum of a function. However this is not so due to
deviations from the equilibrium induced by surface wave.
To calculate $\partial F/\partial Q$ one should take into account the changes
in surface area (and surface pressure) due to dilational wave. This
means that the position of a minimum in the free energy and the corresponding
order parameter Q$_0$ also would change
upon surface compression (rarefaction) due to a strong pressure
dependence of A. Therefore we can take the Q$_0$ derivative, having in mind
that the position of minimum shifts with the change in A: $\partial F/\partial
Q_0=-A Q_0$. This derivative is negative (Q$_0 >0$, A$<0$ in nematic phase)
and the minimum of the free energy is lowered upon an increase in Q$_0$.
Hence the Q derivative of the tension is positive:
\begin{equation}
\label{deriv}
\partial\gamma/\partial Q=\rho_s/M_m N_A k_B T\mid A\mid Q_0
\end{equation}

It turns out that the derivative $\partial\gamma/\partial Q$ is 
positive for the whole Q-range and the surface viscosity is also
positive according to Eq(\ref{mod}) where $\partial Q/\partial
\rho_s <$0 . Searching additional physical arguments in favor 
of an {\it increasing} $\gamma$(Q) dependence one can return to 
Eq(\ref{mod}) introducing surface elasticity in the framework
of orientational model. We may perturb a
film by a small horizontal displacement which rarefies 
locally  the monolayer and decreases the surface
concentration. At the same time the surface expansion would
increase the surface order parameter due to the orientational
effect of the dilational wave discussed above. 
Negative sign in the definition of $\epsilon$ reflects the
feed-back reaction of a perturbed surface element. For instance,
one may expect the surface tension to decrease on expansion of
the surface which is accompanied by some increase in Q. In this
case a restoring force would act outward i.e. from the surface
region with lower $\gamma$ to those regions having higher
tension. Thus a {\it decreasing} $\gamma$(Q) dependence would 
lead to a semi-infinitive expansion and the film instability. On
the contrast, having supposed an increasing $\gamma$(Q) 
dependence the surrounding film would tend to compress the
surface element which was expanded initially. 
This ensures that {\it increasing} $\gamma$(Q) provides the monolayer 
stability setting up $\partial\gamma/\partial Q >$0 and positive
$\epsilon'$. 

The modulus of dilational viscosity given by $\epsilon'=\Im [\epsilon]
/\omega$, found from Eqs. (\ref{mod}), (\ref{firstder}), (\ref{deriv})
\begin{equation}
\epsilon'(\omega)= \frac{k_B TN_A\rho_s\mid A\mid Q_0}
{\mid\omega\mid M_m}
\Im[\frac{\imath\tau}{\imath\tau+2/\omega(A+6BQ_0^2)}]
\end{equation}
 is plotted in Fig 2 for different time
constants $\tau$ using a typical value of the surface 
concentration $\rho_s \approx$ 1 mg/m$^2$. 
This plot ensures that the dilational viscosity exhibits a strong
$\omega$ dependence for some $\tau > 10^{-5}$ s while for shorter time
constants $\epsilon'$ is nearly independent of $\omega$.

Surface viscoelastic
properties of glycerol mono-oleate monolayers on water were studied
by surface light scattering in \cite{paris}. Surface properties were deduced
from a many parametrical fit to the measured autocorrelation functions.
Although the surface viscosity was introduced in the framework of
 a phenomenological model (see above) $\epsilon'\approx 10^-4$ mN s/m
 was found. The dilational viscosity inferred from the fit was sensitive
 to the monolayer compression: the increase in $\epsilon'$ about an order
of magnitude upon the monolayer compression have been reported. 

Our analysis shows that the dilational surface viscosity appears
in the form of oscillations (with some relaxation time) of the order parameter
which take place due to the passage of the dilational wave.
This surface anisotropy provides a phase shift between
the velocity gradient and the in-plane director. 
The surface dilational viscosity introduced in our model is  
positive and drops to zero at the point where A$\geq$0 and Q=0 
i.e a point of the surface phase transition of a second order.
 The main assumption of the present paper is the existence of 
nematic ordering in the 
liquid-expanded state. However the present model of surface 
viscosity can be modified for 
hexatic ordering in Langmuir monolayers.         \\  

Fruitful discussions were shared with E.I. Kats.

\pagebreak
\section*{Captions}
\begin{itemize}

\item[Fig.1]
Orientational effect of the velocity field in the surface
wave on the cross section of a molecule in Langmuir monolayer. q 
is the wavevector of the
dilational wave. The dashed lines show the lines of equal phase
of velocity in longitudinal wave. Surface velocity remains
constant along the lines but varies perpendicularly to them giving rise
to the gradient  du$_x$/dx on the surface. 

\item[Fig.2]
Surface dilational viscosity $\epsilon'(\omega)$
calculated just near the surface phase transition for different relaxations times $\tau$.
$\tau$ are labeled near each curve in s. The following values of
the parameters are used: Q=0.1, $\mid A\mid$=0.4, B=1 

\end{itemize}
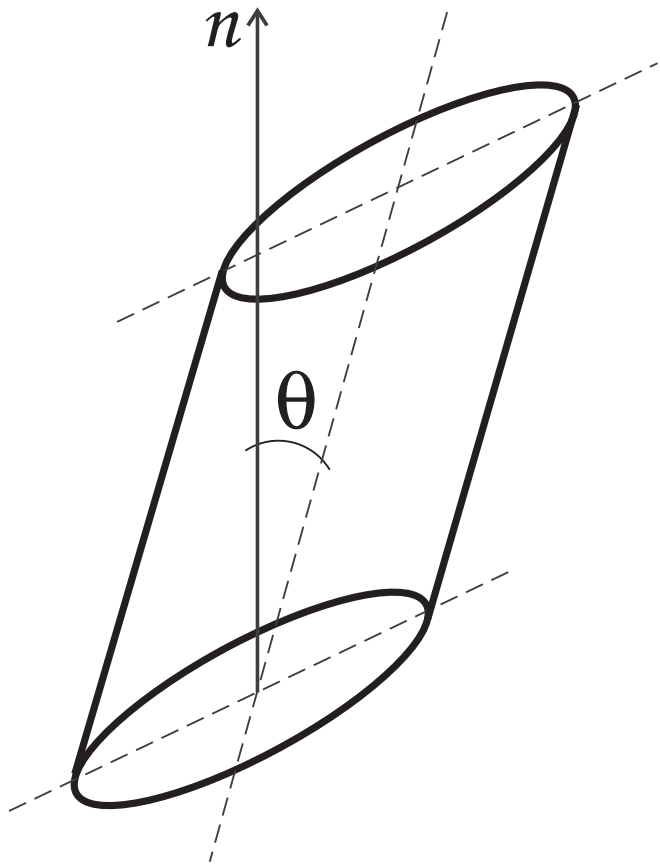
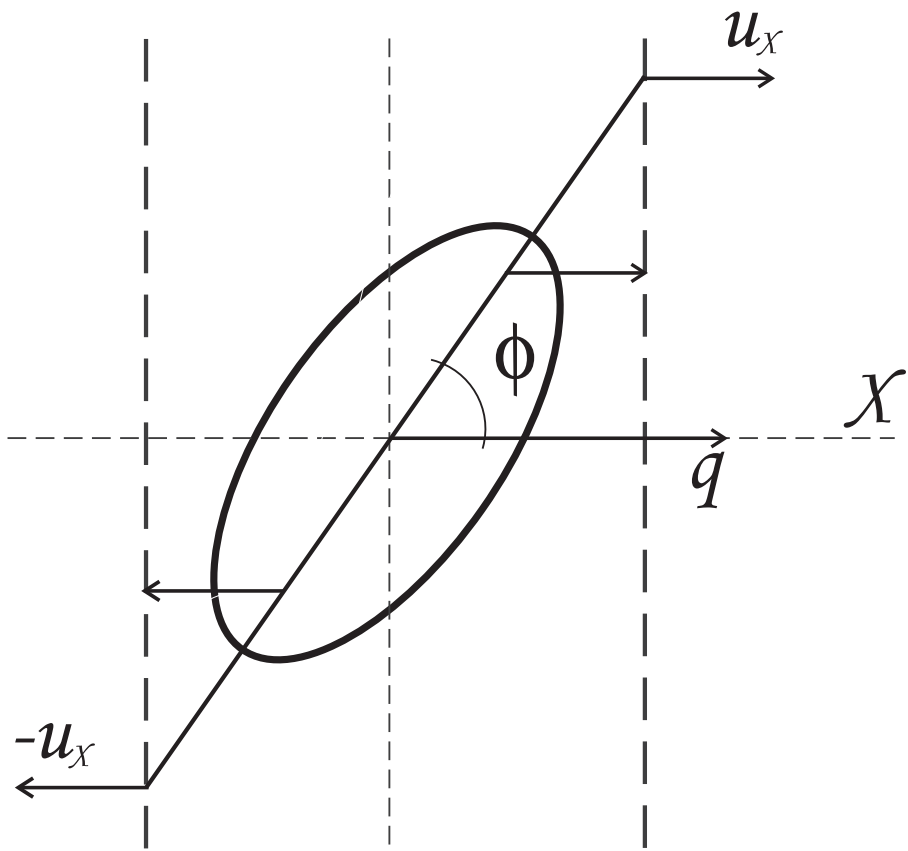
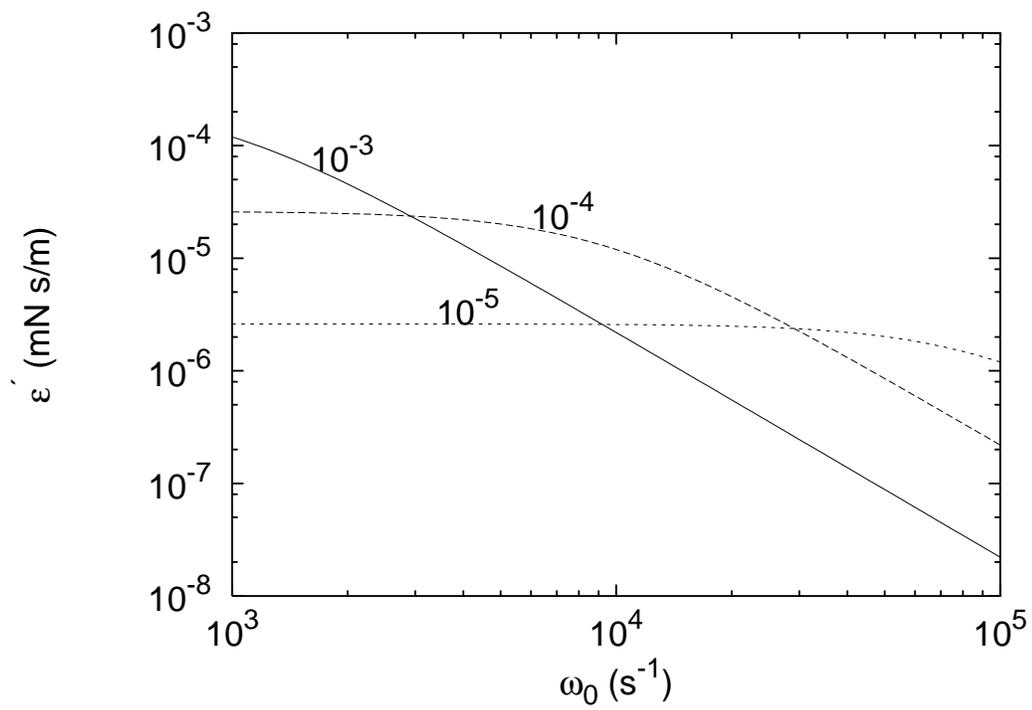
\end{document}